\documentclass[
reprint,
showkeys,
 amsmath,amssymb,
 aps,
]{revtex4-2}

\usepackage{graphicx}
\usepackage{dcolumn}
\usepackage{bm}
\usepackage{amssymb}
\usepackage{amsmath, amsthm}
\usepackage{mathrsfs}
\usepackage{supertabular, multirow}
\usepackage{graphicx}
\usepackage{epsfig}
\usepackage{subfigure}
\usepackage{bm}
\usepackage{xcolor}
\usepackage{algorithm}
\usepackage{tabularx}

\usepackage{setspace}
\usepackage{natbib}

\newtheorem*{theorem*}{Theorem}

\begin{document}

\title{Advanced representation learning for flow field analysis and reconstruction}

\author{Yikai Wang, Jiameng Wang, Ruyi Han, Shujun Fu}
\email[Author to whom all correspondence should be addressed: ]{shujunfu@163.com}
\affiliation{%
School of Matahematics, Shandong University, Jinan 250100, China
}%

\date{\today}

\begin{abstract}
In this paper we present advanced representation learning study on integrating deep learning techniques and sparse approximation, including diffusion models, for advanced flow field analysis and reconstruction. Key applications include super-resolution flow field reconstruction, flow field inpainting, fluid-structure interaction, transient and internal flow analyses, and reduced-order modeling. The study introduces two novel methods: flow diffusions for super-resolution tasks and a sparsity-boosted low-rank model for flow field inpainting. By leveraging cutting-edge methodologies in computational fluid dynamics (CFD), the proposed approaches improve accuracy, computational efficiency, and adaptability, offering deeper insights into complex flow dynamics.
\end{abstract}

\maketitle


\section{Introduction}

In the field of fluid dynamics, obtaining high-fidelity and high-resolution flow field data is crucial for understanding complex fluid behaviors, with applications spanning aerospace, energy, and biomedical engineering. However, experimental and numerical simulations often yield low-resolution data, which limits the ability to capture detailed flow structures, especially in turbulent flows. Traditional computational methods such as Navier-Stokes-based solvers face challenges in computational cost and scalability, especially for high-fidelity simulations \cite{eivazi2022physics, thuerey2020deep}. To address this challenge, researchers have explored various techniques to enhance the resolution of flow field data. Traditional methods, such as interpolation, often fail to accurately reconstruct the intricate details of flow fields. Recent advancements in deep learning offer promising solutions, particularly through the use of diffusion models and convolutional neural networks (CNNs) \cite{kingma2013auto, liu2015deep} and present transformative opportunities to overcome these limitations \cite{kim2019deep, bhatnagar2019prediction, guo2016convolutional}.

This paper focuses on several interrelated topics based on flow field analysis and reconstruction:

Flow field super-resolution reconstruction. Super-resolution reconstruction involves enhancing the spatial resolution of flow data while preserving physical fidelity. Traditional interpolation methods often fail to capture intricate flow structures. Machine learning (ML) and deep learning (DL) models, including CNNs and generative adversarial networks (GANs) \cite{goodfellow2014generative}, have shown remarkable potential in this area \cite{dong2016image}.

Flow field inpainting. Flow field inpainting addresses the problem of missing or corrupted data. Techniques such as physics-informed neural networks (PINNs) \cite{raissi2019physics} enable reconstruction while adhering to governing equations like continuity and Navier-Stokes laws.

Fluid-structure interaction (FSI). Coupled FSI systems require simultaneous resolution of solid deformation and fluid motion. DL techniques are being employed to accelerate iterative coupling processes and predict structural responses in real time.

Transient and internal flow analyses. They focus on studying time-dependent dynamic characteristics of fluid flow and internal behaviors. These analyses are widely applied in engineering and natural phenomena, aiming to enhance the understanding and prediction of complex flow dynamics.

Reduced-order modeling (ROM) and fast computing. Reduced-order models are essential for approximating high-dimensional systems in a computationally efficient manner. DL-based ROMs leverage latent space representations to approximate flow dynamics.

By leveraging ML and DL, we aim to enhance both predictive accuracy and computational efficiency in complex flow analyses. 

\section{Flow field super-resolution reconstruction based on diffusion models}

\begin{figure*}[htbp]
    \centering
    \includegraphics[width=0.6\textwidth]{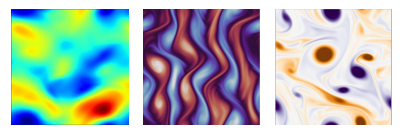}
    \caption{Original flow data for experiments.}\label{ori}
\end{figure*}

In fluid dynamics, obtaining high-fidelity and high-resolution flow field data is crucial for understanding complex fluid behaviors (see Figure \ref{ori} for original flow data) \cite{fukami2023super,shu2023physics,fukami2019super}. Traditional methods like interpolation often fail to accurately reconstruct intricate details of flow fields. Deep learning, particularly diffusion models and CNNs, offers promising solutions. Super-resolution reconstruction in flow fields involves enhancing spatial resolution while preserving physical fidelity. DL models like CNNs and GANs have shown potential, but often lack physical constraints, leading to inaccuracies.

Convolutional Neural Networks (CNNs), known for their spatial feature extraction capabilities, are applied for high-resolution flow reconstructions and transient analyses. Diffusion models are generative techniques that progressively add noise to data and then learn to reverse the noise process. These models are particularly effective for stochastic flow field reconstructions. Physics-Informed Neural Networks (PINNs) integrate physical laws into the loss function, ensuring physically consistent predictions, even in sparse data environments.

Researchers use a multi-scale attention mechanism to capture global and local flow features in flow field super-resolution. The model is trained on flow datasets, focusing on maintaining turbulence characteristics. To reconstruct incomplete flow fields in flow field inpainting, one utilizes GANs conditioned on existing data. Physics constraints are incorporated to maintain continuity and energy conservation. Transient flow fields are analyzed using sequence-to-sequence models, such as recurrent neural networks (RNNs) and attention mechanisms, for time-dependent predictions.

Diffusion models, a type of generative model, have been adapted for flow field reconstruction by modeling the process of adding and removing noise in a flow field. Diffusion models gradually corrupt data with noise and then learn to reverse this process. The training involves a forward diffusion process and a reverse denoising process, using a parameterized Markov chain and a neural network to approximate the conditional distribution.

\subsection{Diffusion model with mixed noise form for flow field super-resolution reconstruction}

Diffusion model for image super-resolution is an image processing technique, which aims to transform low-resolution images into high-resolution ones, so as to enhance the details and clarity of images. Its core lies in the use of self-similarity information of images. In low-resolution images, there are many repeated textures, structures and characteristics, which can be used to infer the missing high-frequency details and generate high-quality high-resolution images. Diffusion model for image super-resolution has a wide range of application prospects. In the fields of medical imaging, satellite remote sensing, video monitoring, etc., this technology can improve the resolution and clarity of images, which is helpful for doctors to diagnose, environmental monitoring and safety monitoring tasks. 

Although the diffusion model for image super-resolution performs well in improving the image quality, it also has some challenges and limitations. For example, the reconstruction results of diffusion generation model have the characteristics of high perception, but it cannot ensure the reduction of fidelity, which makes missing the edge information of super-resolution reconstruction of image, so it is essential to do a good job in the balance of perception and fidelity. In addition, the flow data image has simple structure, less texture information, and one should pay more attention to the reconstruction of edge. Based on this, we propose two innovative points to solve the above problems: 

1. Before super-resolution reconstruction, we add a blind denoising module, which first performs coarse-grained fidelity reduction of flow morphology, and then sends the image into the super-resolution model for reconstruction, which makes the model better balance 
perception with fidelity. 

2. A hybrid diffusion scheme is proposed. Flow images are different from natural images. It lacks texture detail information and focuses more on edge morphology reconstruction. Therefore, longer diffusion steps will bring unnecessary detail reconstruction. Therefore, we shorten the diffusion steps and control the noise intensity at the same 
time. Among them, the noise form is a mixed form of residual and random Gaussian noise, which can improve the robustness of the model while ensuring the reconstruction effect.

\begin{figure*}[htbp]
    \centering
    \includegraphics[width=\textwidth]{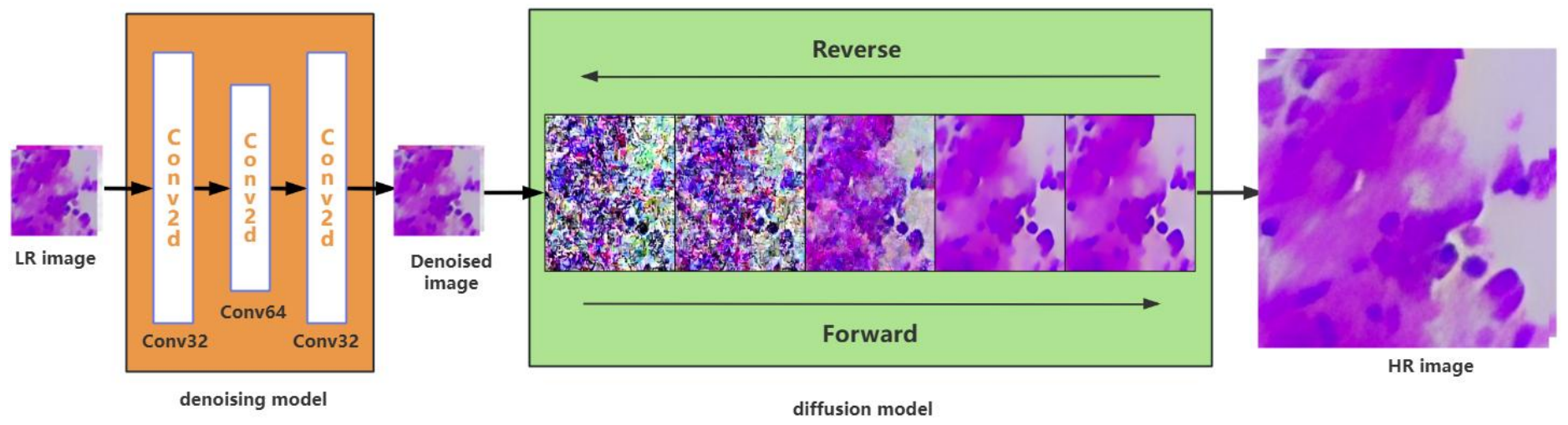}
    \caption{Flowchart for diffusion model with mixed noise form.}\label{df1}
\end{figure*}

The overall process can be divided into two steps (see Figure \ref{df1}):

1. The low-resolution image first enters the denoising module, which is based on Noise2Noise model and uses the characteristics of noise to noise. The module can realize blind denoising using three-layer convolution and ensure the time cost while ensuring coarse-grained denoising. This step can be regarded as the pretreatment process of super-resolution reconstruction, and the main purpose is to improve the fidelity reduction of the model to the original low-resolution image. 

2. The denoised image obtained in the first step is fed into the diffusion model for super-resolution reconstruction. In order to ensure the effect, a 15-step diffusion in the original Resshift model is changed to 6-step to prevent the generation of too many unnecessary details and change the noise strategy. We choose the method of mixing residual errors with random Gaussian noise to add noise, and it turns out that better results can be obtained.

\begin{figure*}[htbp]
    \centering
    \includegraphics[width=\textwidth]{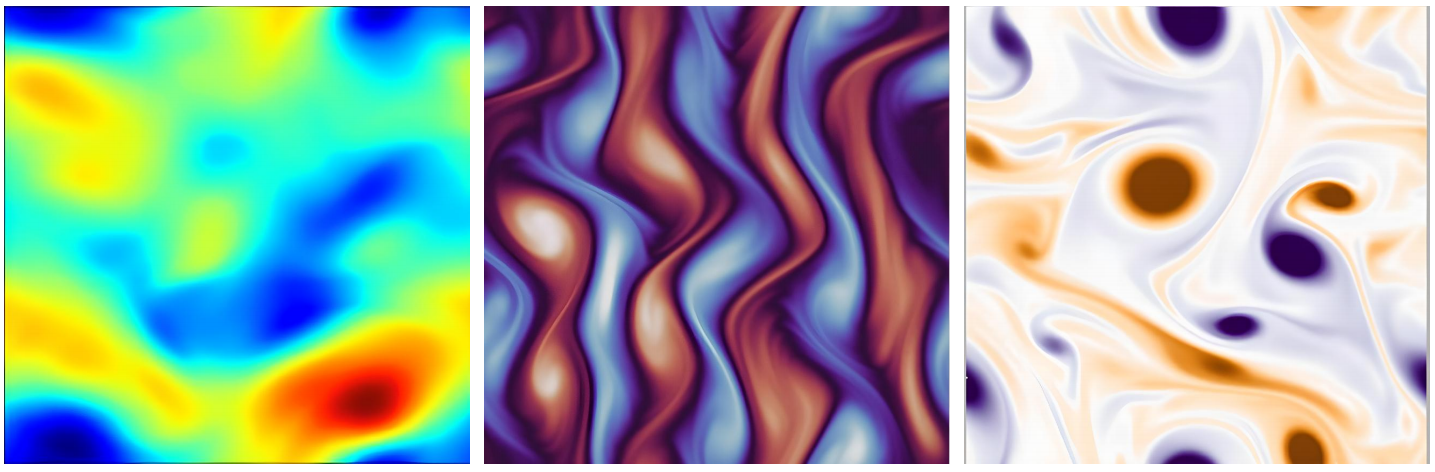}
    \caption{Results of diffusion model with mixed noise form.}\label{redf1}
\end{figure*}

This flow diffusion model demonstrates superior performance compared to traditional interpolation methods and other DL approaches like SRCNN and SRGAN. It achieves high SSIM values for complex turbulent flow fields, even at high downsample factors, showcasing its ability to reconstruct intricate flow features. The model's ability to predict flow field residuals contributes to enhanced performance, focusing on differences between LR and HR flow fields. The use of  fast solver for sampling during inference significantly speeds up reconstruction, making the model suitable for real-time applications (see Figure \ref{redf1}).
    
\subsection{Diffusion model with recursive generalized attention for flow field super-resolution reconstruction}

Single image super-resolution (SR) is a classic problem in computer vision and image processing. It aims to reconstruct a high-resolution image from a given low resolution input. 
In practical situations, many low resolution images cannot be restored by traditional methods due to severe degradation, so deep neural networks have become one of the 
optimal solutions for achieving super-resolution. With the application of Transformer in the image field, it has also demonstrated excellent performance in the super-resolution 
field. However, the transformer itself also faces problems such as computational complexity. Although it becomes linear complexity through window self attention, it loses 
global information and cannot fully utilize image pixels for high-resolution reconstruction, especially in high-frequency information such as microscopic image edge texture, the 
restoration effect is not satisfactory. This article has made some improvements to this section.This method uses a transformer as the infrastructure, and its performance in image 
processing is mainly affected by computational complexity. By using a window self attention module, the computational complexity is reduced, but at the same time, global 
information is lost, and only a small number of pixels participate in image restoration. The method proposed in this paper is mainly aimed at better utilizing global and local 
information, activating more effective pixels to participate in reconstruction. 

\begin{figure*}[htbp]
    \centering
    \includegraphics[width=0.7\textwidth]{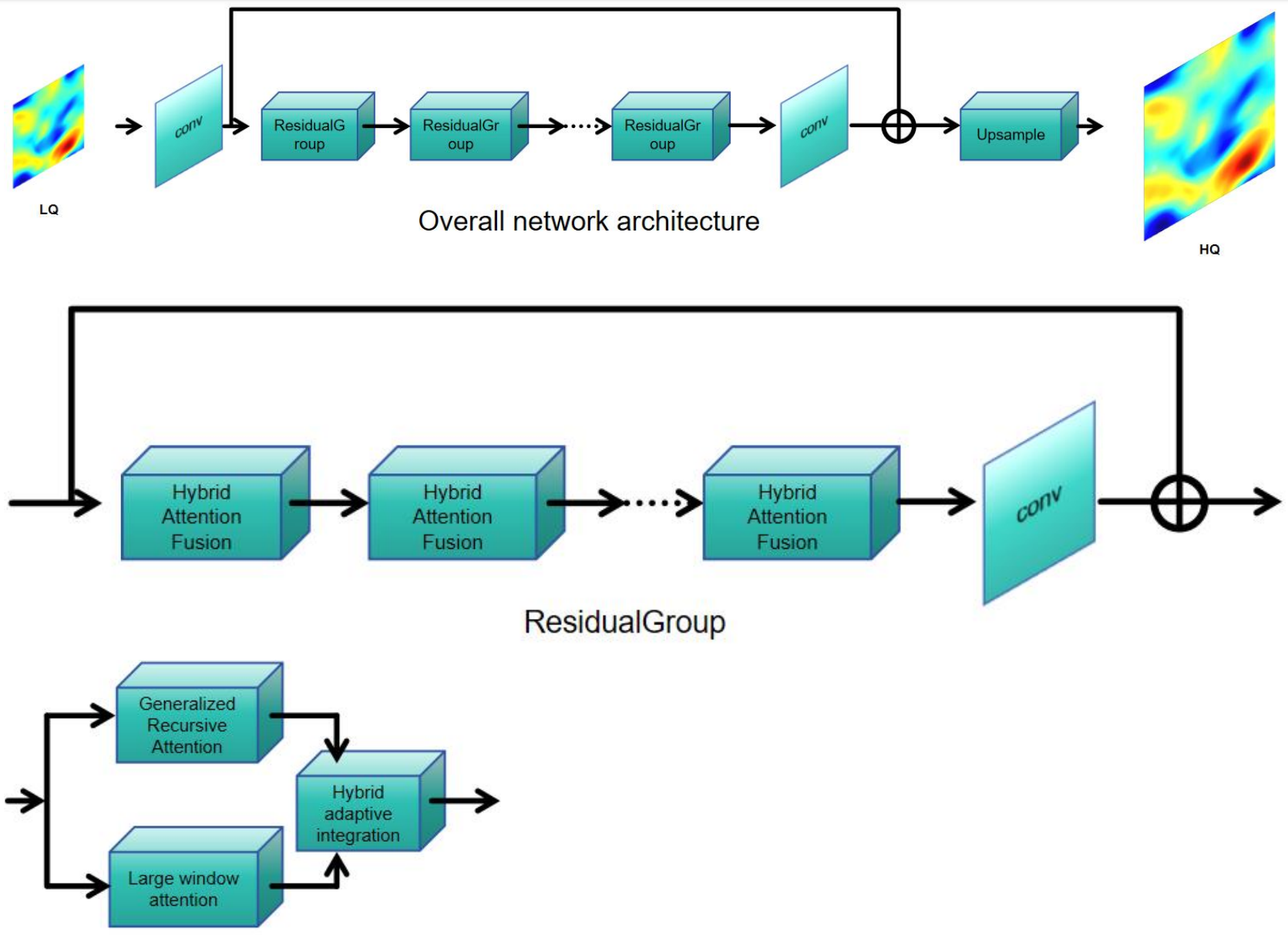}
    \caption{Flowchart for diffusion model with recursive generalized attention.}\label{df2}
\end{figure*}

The main innovations of this method are as follows (see Figure \ref{df2}): 

Firstly, by introducing a recursive generalized attention module to compensate for the lack of global information in the window self attention module; 

Secondly, by optimizing the window self attention module and increasing the window size, more pixels can participate in the reconstruction process; 

Thirdly, by introducing a gating mechanism in the feed forward neural network part, and introducing multi-scale convolution and wavelet information as guidance in the original gating module part, the model's ability to extract high-frequency information such as low resolution image boundaries is improved; 

Fourthly, a new loss function is introduced in the training process, which reduces the artifact phenomenon in the model predicted image by comparing the L1 loss of the model predicted image and the real image at low resolution. 

At the same time, by fine-tuning the generative large model, the training dataset was expanded, which also improved the model performance to a certain extent. 

Specific process is described as follows. After the low resolution image is input into the model, it is separately input into the recursive generalization self attention module to extract global information and the window self attention module to extract local information. The feed forward neural network in the two modules uses an improved gating mechanism to activate the effective pixels in the feature map output by the previous attention mechanism and suppress irrelevant pixels. Through this process, further screening is carried out. The two parts of information are input into the hybrid adaptive integration module for adaptive information fusion. After residual calculation with images that have not been input into this module, they are input into the next attention module. After multiple iterations, the feature map is upsampled to obtain a high-resolution image and mapped to the RGB space to obtain the predicted image from the final model.

\begin{figure*}[htbp]
    \centering
    \includegraphics[width=\textwidth]{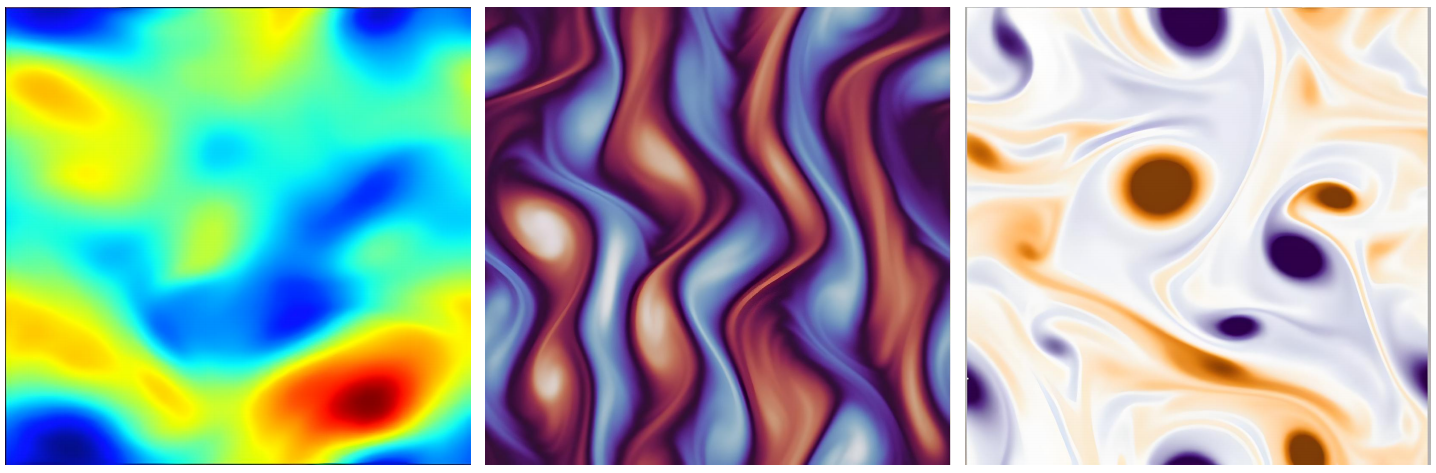}
    \caption{Results of diffusion model with recursive generalized attention.}\label{redf2}
\end{figure*}

The flow diffusion model also outperforms traditional interpolation and some DL methods, achieving high index for complex turbulent flows. Its framework enhances accuracy with desired further applications (see Figure \ref{redf2}).

\section{Transformed sparsity-boosted low-rank model for flow field data completion}

\begin{figure*}[htbp]
    \centering
    \includegraphics[width=\textwidth]{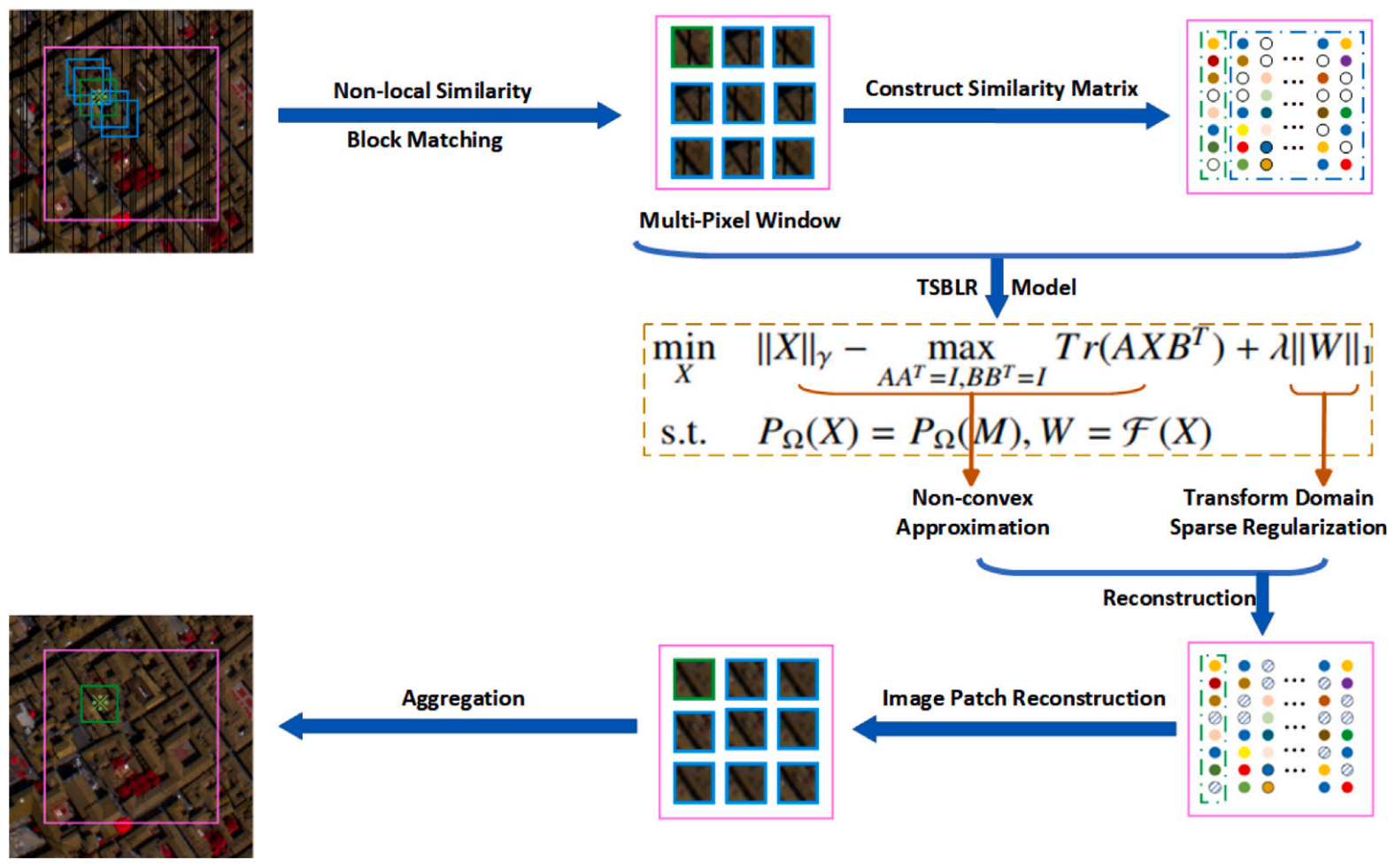}
    \caption{Flowchart for transformed sparsity-boosted low-rank (TSBLR) model.}\label{lr1}
\end{figure*}

Image restoration, particularly image inpainting, is crucial in various fields \cite{zhang2017beyond}. Traditional methods rely on the low-rank property of images but often fail to preserve fine details and edges. This paper introduces the transformed sparsity-boosted low-rank (TSBLR) model \cite{han2025transformed}, incorporating non-convex gamma-norm regularization and non-local prior to enhance image inpainting accuracy and quality (see Figure \ref{lr1}).

Matrix completion techniques are widely used for image restoration. The nuclear norm, as a convex relaxation of the rank function, is popular but limits flexibility. Non-convex rank approximations like the truncated nuclear norm (TNN) offer better control over target rank. The gamma-norm provides a more accurate approximation of the rank function than the nuclear norm. It is defined as a function of the singular values of the matrix, with a parameter gamma controlling the approximation behavior. Replacing the nuclear norm with the gamma-norm in the optimization problem leads to a more precise low-rank approximation.

Transform domain sparse regularization exploits the sparsity of natural images in transform domains like the discrete cosine transform (DCT). By adding a sparsity constraint, the model better reconstructs fine details and textures, resulting in higher quality inpainting. The non-local prior leverages the self-similarity of natural images, searching for similar patches in the neighborhood of the target block to guide the inpainting process. This approach is effective in restoring consistent textures and patterns across the image.

The optimization problem is solved using an alternating direction method of multipliers (ADMM) framework, decomposing the problem into a series of sub-problems. The ADMM algorithm iteratively updates the variables involved, ensuring convergence  (see Algorithm 1).

\begin{figure*}[htbp]
    \centering
    \includegraphics[width=\textwidth]{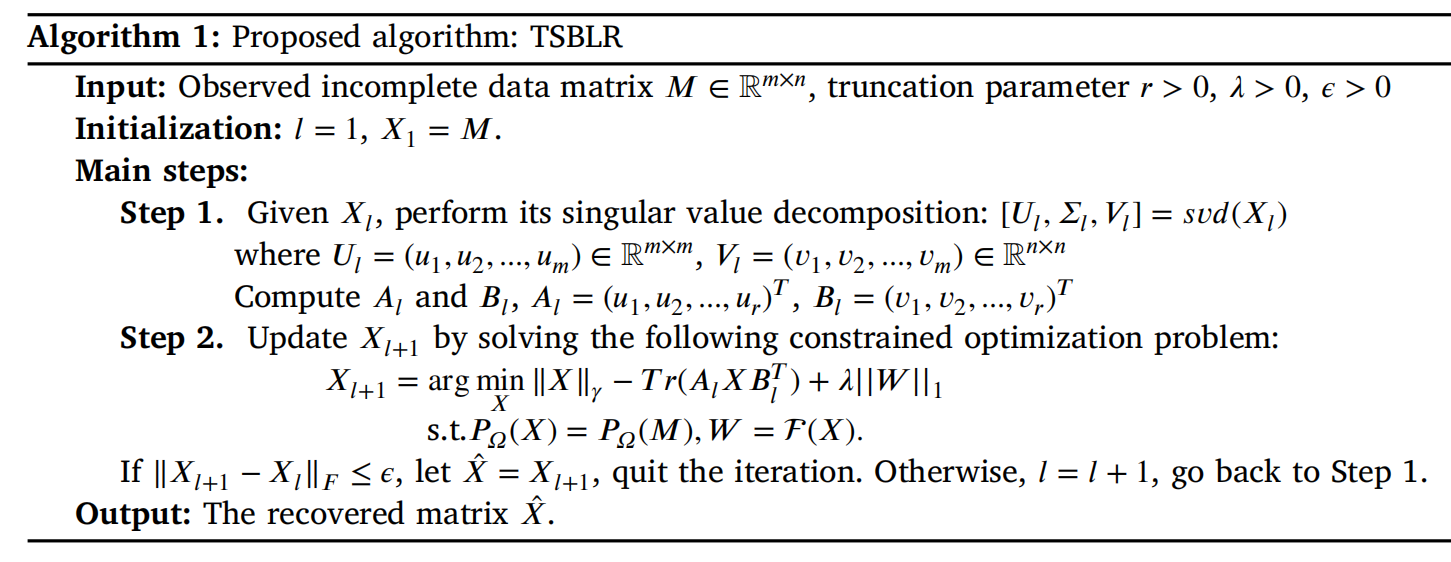}
\end{figure*}

\begin{figure*}[htbp]
    \centering
    \includegraphics[width=0.9\textwidth]{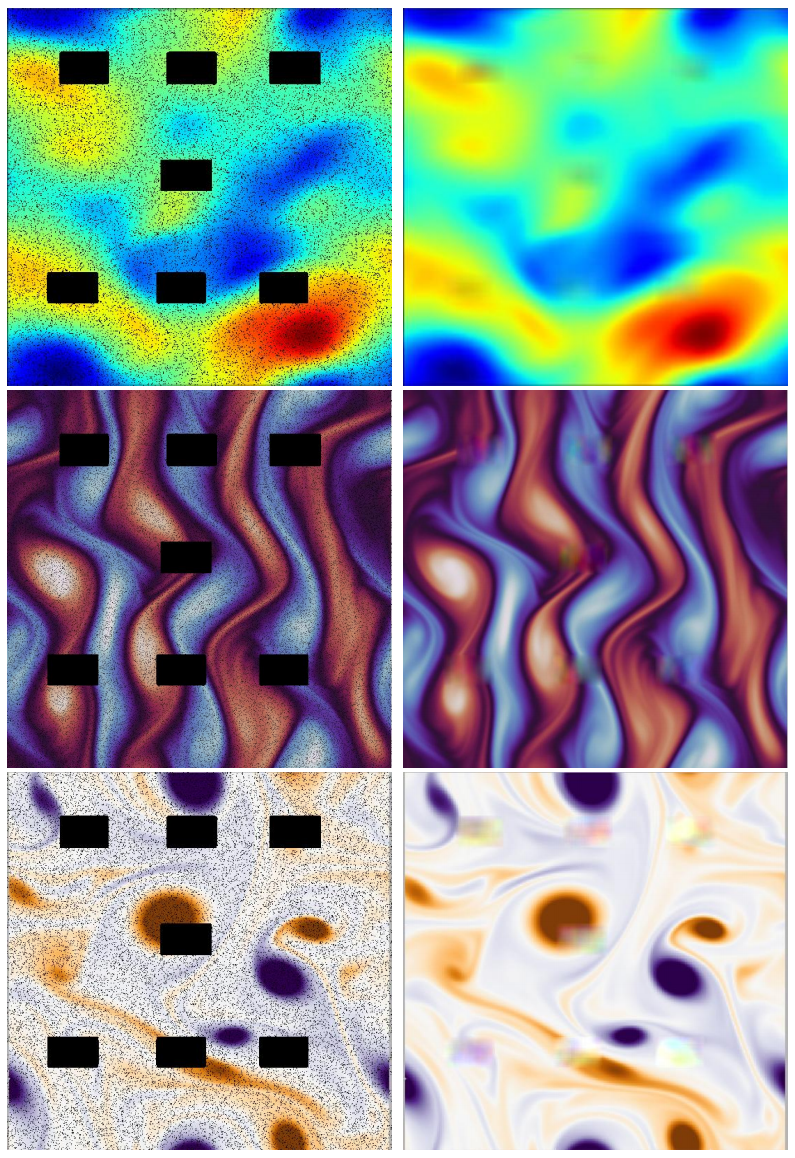}
    \caption{Results of transformed sparsity-boosted low-rank (TSBLR) model.}\label{res}
\end{figure*}

The TSBLR model is evaluated on tasks like pixel inpainting, patch removal and noise removal. It outperforms many low-rank methods in both visual quality and quantitative evaluation. The TSBLR model restores images with sharper edges and more accurate textures, achieving higher PSNR and SSIM values \cite{wang2004image}. It effectively removes loss and restores entire data with fewer artifacts   (see Figure \ref{res}). In the ablation study removing the sparse regularization component leads to a significant decrease in restoration performance, highlighting the importance of the sparse prior in capturing local details and texture information. The model's performance is sensitive to parameters like truncation, penalty, and regularization. Optimal values are determined through extensive experiments to ensure best performance in different scenarios. 

\section{Conclusions}

The paper introduces advanced techniques for flow field analysis and reconstruction, with the flow diffusion model achieving superior super-resolution performance, even for complex turbulent flows, by effectively reconstructing detailed features and enabling real-time applications. Additionally, the transformed sparsity-boosted low-rank model significantly enhances flow field accuracy and quality, restoring images with sharper edges and accurate textures, outperforming traditional methods through its non-convex gamma-norm regularization and non-local prior, which effectively capture local details and minimize artifacts.

\bibliographystyle{IEEEtran}
\bibliography{ref}

\end{document}